\documentstyle[twoside,fleqn,espcrc2]{article}
\def\vev#1{\langle#1\rangle}
\newcommand{\psibar}{\bar\psi}
\newcommand{\cond}{\vev{\psibar\psi}}
\newcommand{\del}{\partial}
\newcommand{\cL}{\cal{L}}
\newcommand{\BG}{{\rm BG}}
\def\half{{\textstyle{1\over2}}}
\def\Tr{\mathop{\rm Tr}\nolimits}

\newcommand{\AmS}{{\protect\the\textfont2
  A\kern-.1667em\lower.5ex\hbox{M}\kern-.125emS}}

\title{Problems with the Quenched Approximation in the Chiral Limit}
\author{Stephen R. Sharpe%
\address{Physics Department, University of Washington, Seattle, WA 98195, USA}%
\thanks{Talk at Lattice '92, Amsterdam, 9/92, to be published in the
proceedings. Preprint UW/PT-92-20, hep-lat/9211005 (November 1992)}}

\begin{document}
\include{psfig}

\begin{abstract}
In the quenched approximation, loops of the light singlet meson
(the $\eta'$) give rise to a type of chiral logarithm absent in full QCD.
These logarithms are singular in the chiral limit
\cite{sslat89,BGpap,sschiral},
throwing doubt upon the utility of the quenched approximation.
In previous work, I summed a class of diagrams,
leading to non-analytic power dependencies such as
$\cond\propto m_q^{-\delta/(1+\delta)}$ \cite{sschiral}.
I suggested, however, that these peculiar results could be redefined away.
Here I give an alternative derivation of the results,
based on the renormalization group,
and argue that they cannot be redefined away.
I discuss the evidence (or lack thereof) for such effects in numerical data.
\end{abstract}

\maketitle

\section{INTRODUCTION}
Many simulations of lattice QCD
use the so-called ``quenched'' approximation,
in which the fermion determinant is left out of the measure.
Since this approximation is likely to be with us for
a few years yet, it behooves us to attempt to understand all we
can about the peculiarities of the quenched theory.

I focus here on the effects of the extra light pseudo-Goldstone
boson (PGB) present in the spectrum of quenched theory.
I assume degenerate light quarks ($N_f$ each of mass $m_q$)
in which case the extra PGB is the flavor singlet state.
By analogy with QCD, I call this the $\eta'$.

The $\eta'$ is a PGB in the quenched theory
because the series of diagrams shown below are absent.
\begin{figure}[h]
\vspace{-0.2truein}
\centerline{\psfig{file=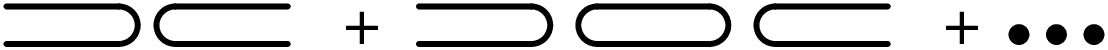,width=2.8truein}}
\vspace{-0.2truein}
\end{figure}
\noindent
The first of these diagrams is, however, present in loop graphs,
and introduces a new dimensionful parameter.
This changes the power counting rules of chiral perturbation
theory, and leads to enhanced chiral logarithms in quenched QCD.

This effect, first pointed out in Ref. \cite{sslat89},
was first calculated by Bernard and Golterman (BG)\cite{BGpap}.
They developed a chiral Lagrangian for quenched QCD, containing
ghost states to cancel the effects of the quark loops.
They found, e.g.
\begin{eqnarray}
\label{BGeqn}
m_\pi^2/2\mu m_q &=& [1 - 2\delta \ln(m_{\eta'}/\Lambda) + O(m_q)] \ , \\
\delta &=&m_0^2/( N_f 16 \pi^2 f_\pi^2 ) \ ,
\end{eqnarray}
where $\Lambda$ is the ultraviolet cutoff,
$f_\pi=93$MeV and $\mu$ are parameters in the chiral Lagrangian,
and $m_0$ is the scale characterizing the $\eta'$ two-point function.
These parameters are discussed further below.
In full QCD, the term proportional to $\delta$ is absent,
and so $m_\pi^2=2 \mu m_q$ in the chiral limit.
Chiral logarithms, proportional to $m_q\ln(m_q)$, vanish in this limit.
(These are represented by the shorthand $O(m_q)$ in Eq. \ref{BGeqn}.)
In quenched QCD, on the other hand, the logarithm diverges
in the chiral limit.

Using the quark-line method I obtained, independently, the same results
for the enhanced chiral logs from $\eta'$ loops \cite{sschiral,ssmorel}.
(I also noted that there are a number of quantities (e.g. $f_\pi$, $B_K$)
which, at one-loop, and for degenerate quarks,
are not affected by $\eta'$ loops.)
In order to address the question of the approach to the chiral limit,
when the term $\delta\ln(m_{\eta'})$ gets large, I summed the leading
logarithms proportional to powers of this quantity, with the result
\begin{equation}
\label{mpieqn}
m_\pi^2/2\mu m_q = (2 \mu m_q/\Lambda^2)^{-\delta/(1+\delta)}
+ O(m_q) \ .
\end{equation}
The divergence has been enhanced from a logarithm to a power law.

This result appears to violate PCAC relations
such as $f_\pi^2 m_\pi^2 = -2 m_q \vev{\psibar\psi}$.
This is not so, however, since there is a whole tower of equations
similar to Eq. \ref{mpieqn},
e.g.
\begin{equation}
\label{condeqn}
-\vev{\psibar\psi}/(\mu f_\pi^2) = (2\mu m_q/\Lambda^2)^{-\delta/(1+\delta)}
+ O(m_q) \ .
\end{equation}

These results are of the same form as those describing critical behavior
close to a second order fixed point.
At first sight this is puzzling, because no such fixed points
(with non-trivial exponents) are known in four dimensions.
The puzzle is resolved by recalling that this lore applies to
critical theories which are unitary,
and thus presumably not to the massless limit of quenched QCD.
The similarity to critical behavior does, however, suggest that
it should be possible to recast the results Eqs. \ref{mpieqn} and
\ref{condeqn} into a renormalization group (RG) framework.

\section{RG DERIVATION}
It is simplest to use the quenched chiral Lagrangian
of Bernard and Golterman, $\cL_\BG$.
The symmetry of $\cL_\BG$ is the graded group $U(3|3)_L\times U(3|3)_R$,
and there are extra pseudoscalar and fermionic PGBs.
It turns out, however, that for amplitudes involving only the physical PGBs
as external particles, and for external momenta $p \ll 4\pi f_\pi$,
one need only consider part of $\cL_\BG$:
\begin{eqnarray}
\nonumber
\cL_\BG &=& \Tr[\del_\mu \pi \del_\mu \pi]
 - \half \mu m_q f_\pi^2 \Tr[U\!+\!U^{\dag}]  \\ &&
\label{clBGeqn}
 \mbox{} + \half m_0^2 \eta'^2 + \cL'_\BG \ .
\end{eqnarray}
The pion field here includes the $\eta'$:
\begin{displaymath}
\pi = \sum_{a=0}^8 \pi_a T_a \ ,
\Tr[T_a T_b] = \half \delta_{ab} \ ,
U = \exp[2 i \pi/f_\pi] \ .
\end{displaymath}

$\cL'_\BG$ includes various types of terms which can be neglected since
they do not give rise to enhanced chiral logarithms.
First, there are other leading order terms (i.e. of $O(p^2,m_q)$)
containing the additional PGB fields, which have the effect of canceling
various diagrams involving the physical PGB's.
These cancellations are discussed in detail in Refs. \cite{BGpap,sschiral}.
Most important here are the restrictions on the insertions of the $\eta'^2$
term. While this looks like a mass term, it can in fact only be inserted
once on each internal $\eta'$ line, because of cancellations with terms
from $\cL'_\BG$.
By contrast, in full QCD $m_0^2$ does contribute to the $\eta'$ mass,
$m_{\eta'}^2 = 2\mu m_q + m_0^2$.
A standard phenomenological analysis using the physical meson masses,
and allowing for non-degenerate quarks,
gives $m_0\approx 0.9$GeV in full QCD.

Using this value as an estimate in the quenched approximation,
one finds $\delta\approx 0.2$.
This estimate is unreliable for two reasons.
First, parameters in the quenched and full chiral Lagrangians
need not be the same.
Second, I have excluded a possible wavefunction renormalization term
$\frac12(A-1)\del_\mu \eta' \del_\mu \eta'$, which reduces the estimate
of $\delta$ by a factor of $A$.
This term does not, however, contribute enhanced chiral logarithms,
because it does not introduce a new scale.
It simply makes the estimate of $\delta$ less certain.

$\cL'_\BG$ also includes the remainder of the full kinetic term
\begin{displaymath}
\frac14 f_\pi^2 \Tr[\del_\mu U \del_\mu U^{\dag}]
= \Tr[\del_\mu \pi \del_\mu \pi] + O(\pi^4) \ .
\end{displaymath}
Only the $O(\pi^2)$ term need be kept, because the higher order vertices
do not contain the $\eta'$ field,
and do not contribute enhanced chiral logarithms.
On the other hand, the full mass term must be kept,
since the $\eta'$ is contained in all terms.

Finally, $\cL'_\BG$ contains $O(p^4)$ terms which
are suppressed in the chiral limit.

I now turn to the calculation.
In full QCD, power counting shows that logarithms always come with
a factor of the quark mass or the external momenta-squared, e.g.
\begin{displaymath}
m_\pi^2/(2\mu m_q) = 1 + 2m_\pi^2 \ln(m_\pi/\Lambda)/(N_f 16 \pi^2
f_\pi^2)\ .
\end{displaymath}
The logarithmic cut-off dependence can thus be absorbed
by the coefficients of the $O(p^4)$ part of the Lagrangian,
while the leading order coefficients are unaffected.

If one uses a simple momentum cut-off, rather than, say, dimensional
regularization, then there are also quadratic divergences.
In full QCD, these do affect the leading order coefficients.
One must introduce a scale dependence in the bare decay constant,
$f(\Lambda)$, so that physical quantities are independent of the cut-off.
These divergences are, however, absent in the quenched theory,
so $f_\pi$ remains independent of scale.


In quenched chiral perturbation theory power counting is changed
because of the additional dimensionful parameter $m_0$.
Insertions of the $\eta'^2$ vertex into diagrams that were previously
quadratically divergent produces logarithmically divergent integrals,
giving factors of $\delta\ln(m_{\eta'}/\Lambda)$. For example,
Fig. 1b leads to Eq. \ref{BGeqn}.
Since $\delta$ does not vanish in the chiral limit,
the scale dependence must be absorbed into the leading order coupling:
$\mu\to \mu(\Lambda)$.
There are also divergent corrections to $m_0^2$,
but these can be shown to be suppressed by powers of
$m_\pi^2/(4\pi f_\pi)^2$.

In Ref. \cite{sschiral} I summed all diagrams containing leading logarithms,
i.e. those proportional to $[\delta\ln(m_{\eta'})]^n$
with no additional factors of $m_\pi^2$ or $\delta$.
Two crucial points allowed this.
First, in order to get a leading logarithm one must insert each $\eta'^2$
vertex in an initially quadratically divergent loop.
Furthermore, every loop in the initial diagram must receive such an
insertion, otherwise the loop will give factors of $m_\pi^2$.
Simple power counting shows that, for degenerate quarks,
the diagrams which remain are those with any number of $\eta'$ loops
coming out of a single mass-term vertex,
as illustrated by Figs. 1b and c. These are easily summed.
Second, other diagrams involving more than one $\mu$ vertex can be shown to
lead to non-leading logarithms using Weinberg's theorem \cite{weinberg}.


\begin{figure}[hb]
\centerline{\psfig{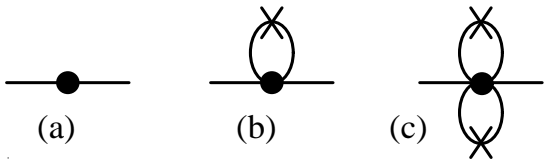}}
\caption{Contributions to $m_\pi$. The cross
represents the $\eta'^2$ vertex, the circle the $\mu$ term.}
\label{fig:largenenough}
\end{figure}

I wish to recast the diagram summation into a RG framework.
Start at a scale $\Lambda_0\approx 1$GeV at which the
coupling is $\mu(\Lambda_0)=\mu_0$.
Reduce $\Lambda$, varying $\mu(\Lambda)$ so as not to change
amplitudes with small external momenta.
When $\Lambda$ drops to $m_{\eta'}$, the loop diagrams cease to
generate logarithms, and $\mu(\Lambda)$ no longer ``runs''.
At this point use the tree level formula to determine $m_{\eta'}$:
\begin{equation}
\label{selfconsistent}
m_\pi^2 = m_{\eta'}^2 = 2 \mu(m_{\eta'}) m_q \ .
\end{equation}

A crucial check on this program is that the
RG equation does not depend on the choice of physical amplitude used to
determine it.
The two-point, four-point, six-point, etc. amplitudes
all yield the same result.
This ensures that $\cL_\BG$ retains its chiral invariant form.

To obtain the RG equation, consider the two point amplitude.
As already noted, the diagrams which contribute at leading order
are those of the type shown in Fig. 1, but with any number of $\eta'$ loops.
Let $A_n$ be the contribution of the diagram with $n$ $\eta'^2$ vertices.
Each loop gives a factor of $\delta\ln(\Lambda^2/m_{\eta'}^2)$,
and there is a combinatoric factor of $1/n!$, so
\begin{displaymath}
A_n = 2 \mu(\Lambda) m_q [\delta\ln(\Lambda^2/m_{\eta'}^2)]^n / n! \ .
\end{displaymath}
Thus if we change $\ln(\Lambda^2)$ by $d\ln(\Lambda^2)$,
\begin{displaymath}
d A_n = \delta A_{n-1} d\ln(\Lambda^2) + A_n d \ln \mu \ ; \ \ A_{-1}=0 \ .
\end{displaymath}
Summing over $n$ we see that the total amplitude
will be unchanged if $\mu(\Lambda)$ changes according to
\begin{equation}
d \ln \mu = - \delta d \ln(\Lambda^2) \Rightarrow
\mu(\Lambda) = \mu_0 (\Lambda_0^2/\Lambda^2)^{\delta} \ .
\end{equation}
The self-consistent equation, Eq. \ref{selfconsistent} becomes
\begin{equation}
m_\pi^2=m_{\eta'}^2 = 2 \mu_0 m_q (\Lambda_0^2/m_{\eta'}^2)^{\delta} \ .
\end{equation}
The solution is Eq. \ref{mpieqn}, except that we now know
that $\mu$ is to be evaluated at the scale $\Lambda$.


More generally, all amplitudes coming from the mass term in $\cL_\BG$,
and involving momenta less than $m_{\eta'}$,
will come with an overall factor of
\begin{equation}
\mu(m_{\eta'}) = \mu_0 (\Lambda_0^2/m_{\eta'}^2)^{\delta} \ .
\end{equation}
Eq. \ref{condeqn} follows from this result.

This derivation is unusual in that one must
work to all orders to obtain the RG equation.
No extra diagrams are summed up when solving the equation.
Furthermore, the external momentum does not act as an infrared cut-off
on the diagram since it does not flow through the $\eta'$ loops.
Nevertheless, I think it is useful to see the result emerge in this way.
For one thing it may simplify the generalization to non-degenerate quarks.
It also emphasizes that the non-analytic dependence on $m_\eta'$ comes from
the infrared part of the integrals.

In Ref. \cite{sschiral} I suggested removing the non-analyticities
by assuming a non-analytic relation between the quark mass in the
quenched chiral Lagrangian and that in quenched QCD.
I no longer think that this is an option.
The two Lagrangians are matched at $\Lambda\sim 1$GeV,
in such a way that their infrared behaviors agree.
Thus infrared divergences leading to non-analyticities
cannot enter into the relationship between quark masses.

\section{COMPARISON TO DATA}

\begin{figure}[tb]
\vspace{2.5truein}
\includegraphics{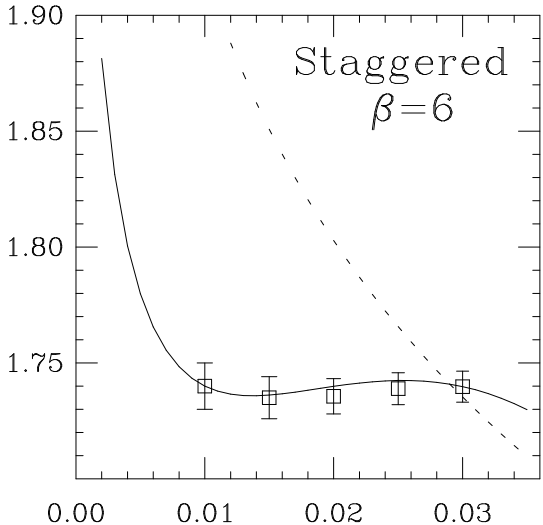}
\caption{$\ln(m_\pi^2/m_q)$ vs. $m_q$ (in lattice units).}
\end{figure}

Present quenched data show no evidence for the divergences of
Eqs. \ref{mpieqn} and \ref{condeqn}.
This is illustrated with the data of Ref. \cite{hmks} in Fig. 2.
Eq. \ref{mpieqn} with $\delta=0.2$ gives the dashed line.
($\Lambda$ is chosen to match the data at $m_q\approx0.03$.)
One is free, however, to add terms linear and quadratic in $m_q$.
With coefficients of typical size,
it is possible to make the theoretical curve quite flat,
as shown by the solid line.
Thus this data set probably cannot rule out $\delta\approx0.2$.
However, new results on the spectrum presented at this conference
are likely to give a much smaller bound on $\delta$.

\section{CONCLUSIONS}

I can think of three explanations for the lack of numerical evidence
for the divergent terms.
(1) The effect could be present (with $\delta\approx 0.2$)
but hidden by a conspiracy of higher order terms.
This I think is unlikely given the latest data,
but I await with interest the results.
(2) The theoretical analysis could be wrong. This seems unlikely,
for the essential physics is in the 1-loop calculation,
and this requires only the presence of the light $\eta'$
and of the $\eta'^2$ vertex.
See also the talk of Golterman \cite{BGlat92}.
(3) The effect is present but $\delta\ll 0.2$. As pointed out to me by
Fukugita, Itoh {\em et al.} attempted to directly measure $m_0^2$
with Wilson fermions \cite{tsukuba}.
They found, with poor statistics, a number $\sim 6$ times smaller
than $(900 {\rm MeV})^2$.

I presently favor the third option.
The only doubt I have stems from the following fact:
to calculate $m_0^2$ one looks at the disconnected part of the
pseudoscalar two point function,
and finds the residue of the double pole at $p^2=-m_\pi^2$.
The same two point function at $p=0$ is proportional to the topological
susceptibility, $\chi$ \cite{smitvink}.
If the double pole gives the dominant contribution to the correlator
at $p=0$ (with no momentum dependence in the residue),
one can show that $f_\pi^2 m_0^2=6\chi$.
Present numerical values of $\chi$ then imply $m_0\approx 900$ MeV.
For $m_0$ to be much smaller than this the residue of the pole must
have considerable momentum dependence.
In the effective Lagrangian this would require a large value of
the coefficient $A$ mentioned above.

If $\delta$ is very small, then the problems with the quenched approximation
are pushed to small masses. This may mean that they can be ignored for
practical purposes.
A direct calculation of $\delta$ would be very useful;
another method for doing this is to study finite volume effects
\cite{BGpap,sschiral}.

\section*{ACKNOWLEDGEMENTS}
I thank C. Bernard, M. Fukugita, M. Golterman, J. Labrenz,
A. Ukawa, and L. Yaffe for useful conversations and comments.
This work was supported by the DOE under contract DE-FG09-91ER40614,
and by an Alfred P. Sloan Fellowship.

%


%

\def\PRL#1#2#3{{Phys. Rev. Lett.} {\bf #1}, #3 (#2) }
\def\PRD#1#2#3{{Phys. Rev.} {\bf D#1}, #3 (#2)}
\def\PLB#1#2#3{{Phys. Lett.} {\bf #1B} (#2) #3}
\def\NPB#1#2#3{{Nucl. Phys.} {\bf B#1} (#2) #3}
\def\NPBPS#1#2#3{{Nucl. Phys.} {\bf B ({Proc. Suppl.}){#1}} (#2) #3}

\end{document}